# Transient evolution of solitary electron holes in low pressure laboratory plasma

Mangilal Choudhary*, Satyananda Kar[a] and Subroto Mukherjee

Institute for Plasma Research, Bhat, Gandhinagar – 382 428, India

[a]Present Address: Department of Mechanical Science and Engineering, Nagoya University, Furo-cho, Chikusa-ku, Nagoya - 464 8603, Japan

*jaiijichoudhary@gmail.com

## ABSTRACT

Solitary electrons holes (SEHs) are localized electrostatic positive potential structures in collisionless plasmas. These are vortex-like structures in the electron phase space. Its existence is cause of distortion of the electron distribution in the resonant region. These are explained theoretically first time by Schamel et.al [*Phys. Scr*. **20,** 336 (1979) and *Phys. Plasmas* **19,** 020501 (2012)]. Propagating solitary electron holes can also be formed in a laboratory plasma when a fast rising high positive voltage pulse is applied to a metallic electrode [Kar et. al., *Phys. Plasmas* **17**, 102113 (2010)] immersed in a low pressure plasma. The temporal evolution of these structures can be studied by measuring the transient electron distribution function (EDF). In the present work, transient EDF is measured after formation of a solitary electron hole in nearly uniform, unmagnetized, and collisionless plasma for applied pulse width $\tau_p < 3\tau_i$ and $\tau_p > 3\tau_i$, where $\tau_p$ and $\tau_i$ are applied pulse width and inverse of ion plasma frequency respectively. For both type of pulse widths, double hump like profile of transient EDF is observed, indicating that solitary electron hole exists in the system for time periods longer than the applied pulse duration. The beam (or free) electrons along with trapped (or bulk) electrons gives the solution of SEHs in the plasma. Without free or beam electrons, no SEHs exist. Transient EDF measurements reveal the existence and evolution of SEHs in the plasma. Measurements show that these structures live in system for longer time in the low pressure range. In high pressure cases, only single hump like transient EDF is observed i.e. only trapped or bulk electrons. In this situation, SEH does not exist in the plasma during evolution of plasma after the end of applied pulse.

## I. INTRODUCTION

Solitary electrons holes are self–consistent plasma structures, associated with positive potential, trap the plasma electrons and form vortices or holes in the phase space. However, depression of electron density in plasma is produced as the hollow vortex structures (SEH) in phase space. These phase space hole is a class of BGK mode [1]. These structures were observed first time as a saturation state of "two stream-instability" in computer simulation [2]. Analytically descriptions of electron hole are explained by Schamel and co-workers [3-9]. Solitary electron holes can be expected whenever $\lambda_D$ << R/5, where $\lambda_D$ is the electron Debye length and $R$ is plasma radius [3]. This condition is satisfied in our experiments. Dynamics and stability of SEHs are described in Ref [10]. Turikov [11] has described the temporal evolution of SEHs by numerical methods. These structures are very sensitive to perturbation of trapped particles orbit via collisional effect (electron–neutral collisions) i.e. with increasing the pressure; these structures will destroy [12]. Kako et.al [13] showed that the electron hole deformation is a slow process. The influences of ion dynamics on the electron hole in phase space have been reported in Ref. 14 and 15. Many satellite observations in Earth's ionosphere and magnetosphere have also reported the existence of solitary pulses with "positive potential spikes". These spikes were nothing but solitary electron hole [16-18]. In space plasma, these structures are extended up to few tens of Debye length parallel to magnetic field. SEHs are often formed as the saturation stage of beam instabilities, transient electrostatics disturbances and magnetic reconnection.

Excitation of SEHs by sudden electrostatics perturbation was observed, first time, in laboratory plasma in Q- machines [12, 19]. For excitation of SEHs, applied potential should be above a critical value $\varphi_c = \frac{m}{2e}(\omega_p a/2.4)^2$ [19], where $m$ is the electron mass, $\omega_p$ is the electron plasma frequency, and $a$ is the plasma column radius. Recently, experimental excitation of SEHs in the laboratory plasma by high voltage electrostatic pulses with time duration $\tau_p > 3\tau_i$ and $\tau_p < 3\tau_i$ is observed with metallic exciter [20] and with dielectric covered metallic exciter [21]. Fox et.al [22] has observed the electron phase- space hole during the magnetic reconnection experiments on the VTF.

Similar to the electron hole, ion holes are also localized electrostatics structures in the ion phase space [5]. Franck et.al [23] observed the periodic ion holes in double plasma device in laboratory. These observed ion holes belong to a class of fully nonlinear solution of Vlasov–Poission's equations of systems. The conditions for validity of Vlasov–Poission description of electron hole or ion hole are one–dimensionality and colissionless plasmas. Both conditions are met in the magnetosphere, fusion machines, particle accelerators, storage rings and are also relevant in laboratory unmagnetized cold plasmas [21, 23]. In low temperature bounded laboratory plasma, for negligible collisions, the given condition (g<<1) where, $g = (4/3\pi n\lambda_D^3)^{-1}$ should be satisfied. In our experiment, $n = 10^{15} m^{-3}$, and $\lambda_D \approx 0.3mm$ and g<<1 is satisfied. Theoretically, first time self -consistent description of solitary electron hole have been carried out by Schamel [3, 5], on the basis of a method to solve the Vlasov-Poission's equations with lower order approximation. For looking the properties of SEHs, velocity limits, potential profile and particle distributions in presence of such structures, our need to get analytical solution of SEHs in 1–D, collisionless, unmagnetized, uniform plasma. Following the Schamel's analytical model for 1–D collisionless, unmagnetized plasma [5, 9] is given. The electron motion in phase space is governed by the Vlasov equation:

$$[\partial_t + v\partial_x + \Phi'(x,t)\partial_v]f(x,v,t) = 0 \tag{1}$$

For stationary electrostatic wave solution, Equation (1) takes the form as:

$$[v\partial_x + \Phi'(x)\partial_v]f(x,v) = 0 \tag{2}$$

An appropriate solution of equation is given in Ref. 9 as

$$f(x,v)=\frac{1+k_0^2\Psi/2}{\sqrt{2\pi}}\left[\left[\theta(\varepsilon)\exp\left(-\frac{1}{2}\left(\sigma\sqrt{2\varepsilon}+v_0\right)^2\right)\right]+\theta(-\varepsilon)\exp\left(\frac{v_0^2}{2}\right)\exp(-\beta\varepsilon)\right] \tag{3}$$

where $\theta(\varepsilon)$ is a Heaviside step function, $\sigma=sg(v)$ is the sign of velocity, $v_0$ is phase velocity of these structures. $\varepsilon=\frac{v^2}{2}-\Phi(x)$ is the single electron energy. Its value depends on $\Phi(x)$ which is an electrostatic potential. Distribution function in Equation (3) has two parts. First part represents the beam (or free) electrons for $\varepsilon\geq 0$ and second part represents the trapped electrons for $\varepsilon<0$. $\varepsilon=0$, is a contour in phase space which separates both trapped and free electron trajectories. This distribution function is continuous in the phase space, i.e., discontinuity is not across the separatrix.

Parameter $\beta$ in eq. (3) is termed as a trapping parameter. This $\beta$ controls the population of trapped particles (electrons). The trapped electron distribution is hole like for $\beta<0$. In absence of trapped electrons, i.e., at $\Phi(x)=0$, it corresponds to the distribution of particles in unperturbed state or shifted Maxwellian in the wave frame.

For finding a self–consistent solution, Poisson equation in the immobile ion limit is used. So Poisson equation can be written as:

$$\Phi''(x)=\int f(x,v)dv-1 \tag{4}$$

Introducing the classical potential (Pseudo–potential) method to find the final shape of potential structure, then eq. (4) can be written in the form [5].

$$\Phi''(x)=\left(-\frac{\partial V}{\partial\Phi}\right) \tag{5}$$

Integration of Eq. (5) gives the Pseudo- energy relation.

$$\frac{1}{2}\Phi'(x)^2+V\left(\Phi;v_{0,}\ \beta\right)=0 \tag{6}$$

Here $V(0;v_0,\beta)=\ 0$ is assumed. Correspondence between the electric potential $\Phi(x)$ and classical potential $V(\Phi;v_0,\beta)$ is given in Ref. 5.

The time independent solution of the Vlasov-Poisson system can be obtained by implying further two conditions on $V(\Phi)$ [5, 9], as

$$V(\Phi;v_0,\beta)<0 \quad \text{In the range} \quad 0<\Phi<\Psi \tag{7}$$

$$V(\Phi=\Psi;v_0,\beta)=0\,. \tag{8}$$

The condition in Eq. (8) gives the nonlinear dispersion relation (NDR). For small amplitude limit, $\Psi << 1$ solution of NDR is given in Ref. 9, which indicates that there exist two regions for the phase velocity of electrostatic structure in the plasma column: a slow velocity branch with $0 < v_{os} \leq 2.13$ and a fast branch with $2.13 \leq v_{of}$ ; $v_{os}, v_{of}$ are normalized velocities of slow and fast branches by electron thermal velocity. Fast branch with $2.13 \leq v_{of}$ represents the classical Langmuir branch modified by particle trapping, and slow branch with $0 < v_{os} \leq 2.13$ represents the phase velocities of structures in the thermal range. Since, SEH is a member of slow branch under some conditions [9] with velocity $0 \leq v_0 < 1.307$. Here $v_0$ is the phase velocity of solitary electron hole normalized by electron thermal velocity.

Initially, plasma is in equilibrium, therefore electrons obey nearly Maxwellian distribution. Since, the loss rate is balanced by the formation rate; therefore plasma is in the stable state. In our experiments, first, we excite the SEHs in the unmagnetized, collisionless, uniform plasma by applying the high voltage positive pulse (above a critical value) to a metallic disk which is assumed as an exciter. We have excited the SEHs for pulse width $\tau_p > 3\tau_i (\tau_p \approx 10\mu s)$ and $\tau_p < 3\tau_i$ $(\tau_p \approx 900 ns)$. Here, $\tau_i$ is the inverse of ion plasma frequency that can be calculated for a given density of plasma. In these experiments $\tau_i \approx 1\mu s$. Propagation of these structures are measured by using another metallic disk probe. Propagation velocity of such structures is given in Ref. 20. Due to SEH formation, plasma goes away from initial equilibrium. Thus, the plasma system will have two types of electron populations (free or beam and trapped); therefore the transient EDF will be double hump like. This transient EDF measurement has not been reported experimentally in laboratory plasmas.

The present report describes the measurement of transient EDF after formation of SEHs in the plasma for longer ($\tau_p > 3\tau_i$) and shorter ($\tau_p < 3\tau_i$) applied pulse widths. From these results we found that in presence of SEHs there exists a beam or free component along with trapped population in the system .i.e., double hump profile of transient EDF. Hence, our experimental results are according to the theoretical prediction of an electron distribution function in presence of SEHs in the plasma system. This paper is organized as: Section 2A describes the experimental setup, Section 2B describes the propagation of SEH and Section 2C describes the procedure for measuring the transient EDF. Results and Discussions are given in Section 3 and conclusions about observed results is given in the Section 4.

## II. EXPERIMENTAL SETUP, 'SEH' PROPAGATION AND 'EDF' MEASUREMENT

**A.** **EXPERIMENTAL SETUP:** -The schematic diagram of the system is shown in Fig. 1(a). The experiments were performed in a grounded cylindrical chamber of stainless steel 304 with an inner diameter of 29 cm and a length of 50 cm. The chamber was evacuated by using a combination of rotary and diffusion pumps and the base pressure was about $5\times10^{-5}$ mbar. The plasma was generated by impact ionization of argon, at pressure of $1\times10^{-3}$ mbar, by primary electrons coming out from dc biased hot thoriated tungsten filaments of diameter 0.25 mm. The filaments were mounted on two stainless steel rings, and biased to a potential of -65 V. The plasma density ($n$) and electron temperature ($T_e$) were determined using a one side - planar Langmuir probe of 4 mm radius for various axial and radial positions. Since, the probe should satisfy the criteria ($\lambda_m >> r_p >> \lambda_D$, here $\lambda_m, r_p, \lambda_D$ are electron – neutral collision mean free path, probe radius and Debye length respectively). Sheridan [24] has provided the range of radius for planar disk probe ( $10 < \frac{r_P}{\lambda_D} < 45$ ) for quantitative determination of plasma parameters. In our experimental case, this ratio is around 14. Therefore, the planar disk probe (4 mm) will not perturb the global plasma. For a fixed discharge current ($I_d$), $n$ and $T_e$ were found to be nearly uniform throughout the plasma column of about 20 cm from exciter position. The experiments were performed with plasma densities 1–5×10$^{9}$ cm$^{-3}$, electron temperatures $T_e$~0.5-2 eV which was measured by the Langmuir probe [25, 26 and 27]. The ion temperature was assumed to $T_i$ ~ 0.05eV.

The pulse-forming circuit is described in Fig. 1(b). For the experimental requirement the values of the capacitor ($C$) and the resistor ($R_2$) were adjusted in such a way that the pulse duration always satisfied either of the conditions $\tau_p < 3\tau_i$ and $\tau_p > 3\tau_i$ and the pulse height was decided by the voltage to which capacitor was charged. The value of the inductor ($L$) was fixed.

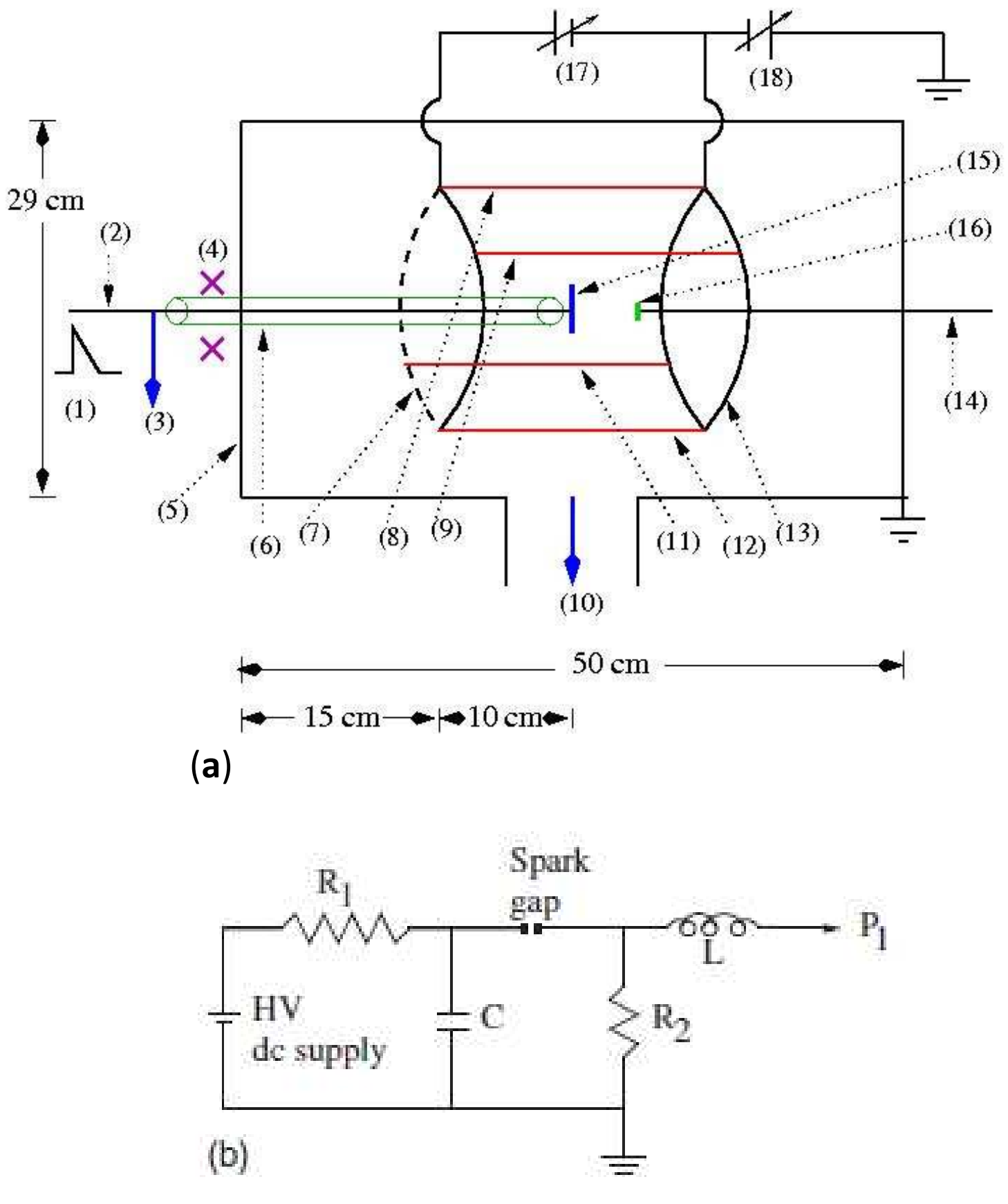

**Fig. 1. (a)** Schematic of the experimental set-up:- (1) Positive pulse to exciter. (2). and (14) are stainless steel rods. (3) High voltage probe to the applied potential to the exciter. (4) Current transformer to measure the current. (5) Main chamber. (6) Ceramic tube. (7) and (13) are stainless steel rings. (8), (9), (11) and (12) are thoriated tungsten filaments. (10) Pumping assembly. (15) Metallic

disc or exciter. (16) Detecting probe or Langmuir probe. (17) Filament heating supply. (18) Discharge power supply. **(b)** Schematic of pulse forming circuit. High voltage dc power supply (1.5 kV, 0.5 A), $R_1$ (40 KΩ) is safety resistance. $R_2$ (100 Ω) is load resistance. *L* (320 nH) is inductance coil. *C* is capacitor and $P_1$ is connected to the exciter plate.

**B.** SEH PROPAGATION - Plasma disturbances were excited by applying a large positive voltage pulse to a metal plate (exciter) inserted in low pressure argon plasma. The pulse magnitudes and durations were in the range of 0.4 – 1.1 kV and 0.14 - 10 μs respectively. In the main chamber, the exciter (stainless steel disc of diameter 10.2 cm and thickness 0.4 cm) was mounted on a stainless steel rod and the stainless steel rod was fixed along the axis of the chamber. A high-voltage probe (Tektronix make 1000X probe) and a current transformer CT (Bergoz make of sensitivity 1 *V/A*) were mounted on the stainless steel rod and were used to measure the pulse voltage and the current respectively. One axially movable probe (detecting probe) of diameter 7 cm and thickness 0.4 cm was introduced to detect the propagation of SEHs. The detecting probe was kept floating and the propagating SEHs were observed directly on an oscilloscope. In Fig .2(a), the applied pulse and corresponding current collected is shown for $\tau_p > 3\tau_i$. The corresponding floating potential variation is given in Fig.2 (b). Fig.3 (a) is showing the applied pulse and corresponding collected current for $\tau_p < 3\tau_i$. Floating potential variation corresponding to $\tau_p < 3\tau_i$ is given in fig 3. (b). In floating potential variation (Figs. 2b & 3b), first peak is corresponding to the electrostatics pick-up ( due to variation in sheath capacitance ) and second one is associated with the propagating positive potential structures (SEHs) in plasma column.

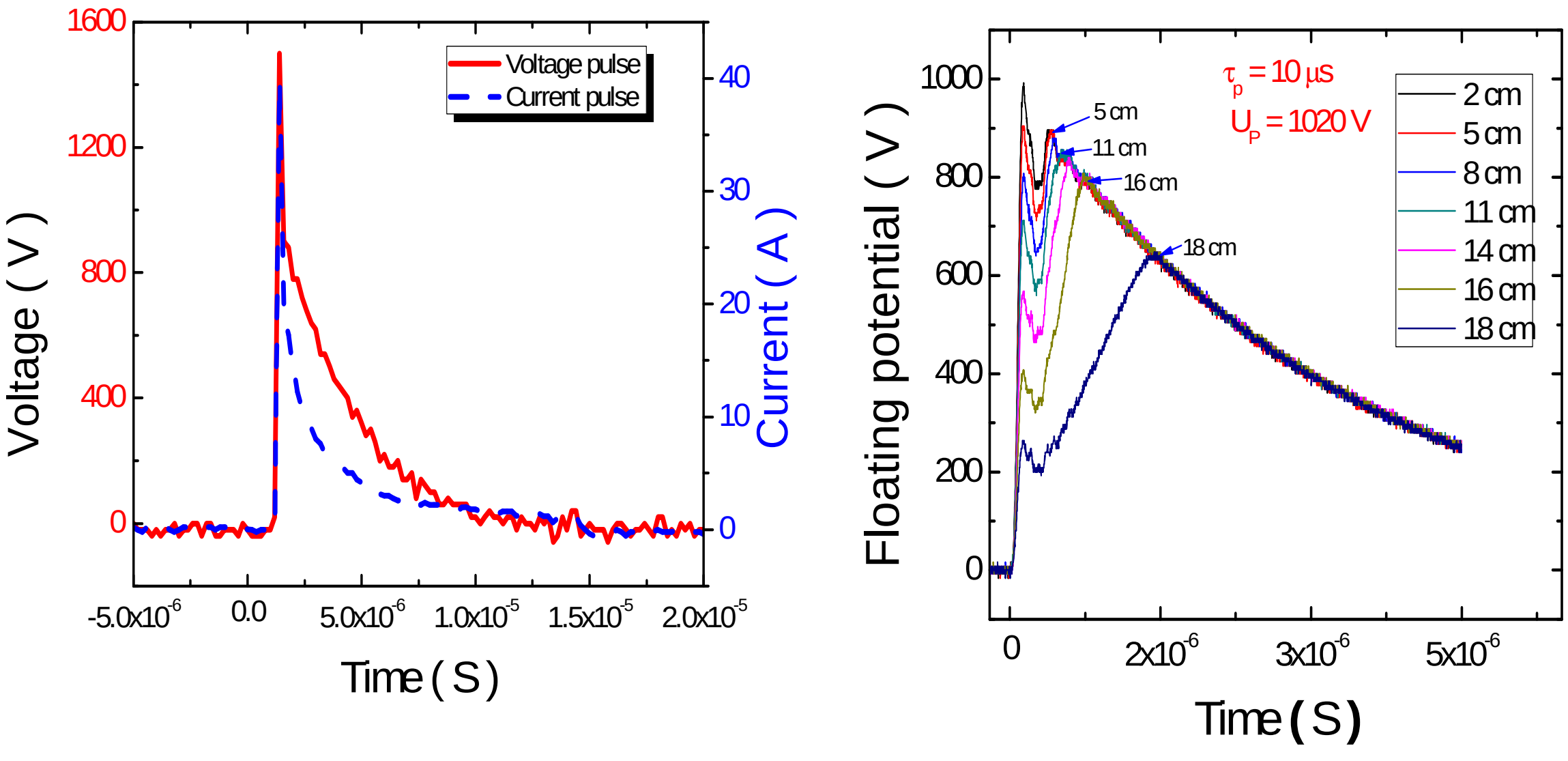

**Fig. 2. (a)**.Applied voltage and corresponding current pulses ($\tau_p > 3\tau_i$). **(b)** Floating potential variations for pulse amplitude 1020 V and pulse width 10 $\mu s$ ($\tau_p > 3\tau_i$) along axial direction from exciter.

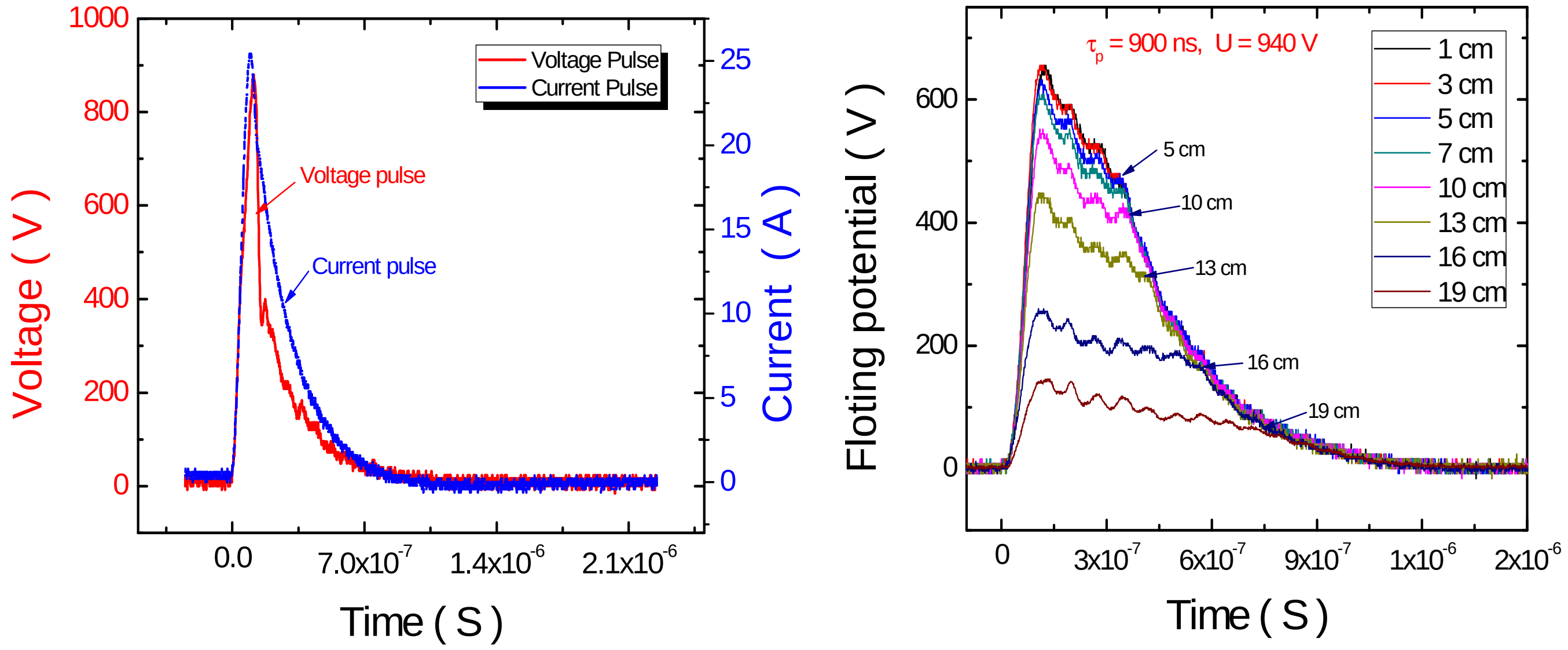

**Fig. 3**. **(a)** Applied voltage and corresponding current pulses ($\tau_p < 3\tau_i$). **(b)** Floating potential variations for pulse amplitude 940 V and pulse width 900 $ns$ ($\tau_p < 3\tau_i$) along axial direction from exciter

**C.** MEASUREMENT OF TRANSIENT EDF: -To measure the transient EDF after excitation of SEHs, a one sided-planar Langmuir probe was used. The Langmuir probe was biased positive (to collect electron current) and negative (to collect ion current) across a 10 k$\Omega$ resistor. This planar probe is used for large collecting area (compare to cylindrical probe), which makes easier to characterize the low density 1–D plasma. The transient EDF is proportional to the first derivative of the planar Langmuir probe *I*-*V* characteristics [28, 29].

Let, an equilibrium electron distribution function is $F_0$. If we perturbed the plasma then distribution function becomes $F_t = F_0 + F_1$, here $F_1$ is perturbed part of EDF at a given time. For looking at the transient part of EDF, we need to measure the time varying part of EDF i.e. $F_1$. After a long time of perturbation, $F_t$ goes to $F_0$.

For measuring the transient EDF in the plasma column, we used one-sided planar probe technique because of less sensitive to noise in the experiments. In our experiments the plane of probe surface is parallel to the exciter surface. It is a 1–D case. Let $V_P$ is probe bias potential and $V_S$ is the plasma potential with respect to a reference potential. Here, potential of chamber wall (reference electrode) is taken as reference point for measurements. For the retardation region, in which $V = V_P - V_S < 0$, the electron current collected by the probe is [28];

$$I_P(V_P) = Ae \int_{\sqrt{\frac{-2e(V_P - V_S)}{m_e}}}^{\infty} v_z dv_z F(v_z) \quad (9)$$

where, $A$ is the area of probe, $m_e$ is the electron mass, $e$ is the electron charge, and $F(v_z)$ is $z$ - projected electron velocity distribution function. In terms of energy scale, collected probe current is given in Ref [29]. Since, the plasma potential is constant for a given time, therefore, integration is written in the given form.

$$I_P(V_P) = \frac{Ae^2}{m_e}\int_0^\infty F(E)dV_P \tag{10}$$

$E = \sqrt{\frac{-2e(V_P - V_s)}{m_e}}$ is energy of a single electron in the plasma. Differentiation of equation (10) with respect to $V_P$ gives the expression as:

$$\frac{dI_P}{dV_P} = \frac{Ae^2}{m_e}F(E) \tag{11}$$

So the EDF is proportional to the first derivative of the planar Langmuir probe *I-V* characteristic with respect to probe bias voltage in the electron retardation region .i.e.

$$F(E) = \frac{m_e}{Ae^2}\frac{dI_P}{dV_P} \tag{12}$$

Now, transient electron distribution function can be expressed as:

$$F_1(E) = \frac{m_e}{Ae^2}\frac{dI'_P}{dV_P} \tag{13}$$

where, $I'_P$ is the temporal perturbed electron current collected by probe. During the excitation of SEHs, transient current to the probe is measured for different probe bias voltages (ion current for negative bias and electron current for positive bias). We have extracted the current for corresponding probe bias for a fixed time and the current–voltage characteristics are obtained for various times. After subtracting the ion saturation current from *I–V* characteristic, we get the perturbing electron current- voltage characteristics for different time after application of pulse. First derivative of the *I–V* characteristic gives the transient EDF at a particular time of pulse application. This measured transient EDF gives the qualitative information about the plasma evolution after the end of applied pulses.

## III. EXPERIMENTAL RESULTS AND DISCUSSIONS:

In unperturbed state, first derivative of *I–V* characteristic is defined as unperturbed or equilibrium EDF. However, the collected current by the probe depends on the probe bias potential with respect to plasma potential. Collected electrons at bias potential near to plasma potential have less kinetic energy but collected electrons have more kinetic energy when the bias potential is more negative to plasma potential. In conclusion, the probe bias potential axis is an equivalent to the energy axis for a constant plasma potential. Therefore, our measurement of EDF is a without defining zero energy point. Since, our objective of experiments is to find the populations present in the plasma at a given time i.e. measured EDF gives the qualitative information of plasma system not quantitative.

Figure 4 shows the EDF with arbitrary unit, in unperturbed state, as a function of probe potential. It is a single hump profile. The shape of EDF is a signature of existence of particles population in the system. Therefore single hump profile indicates one type electron population in the plasma under given conditions.

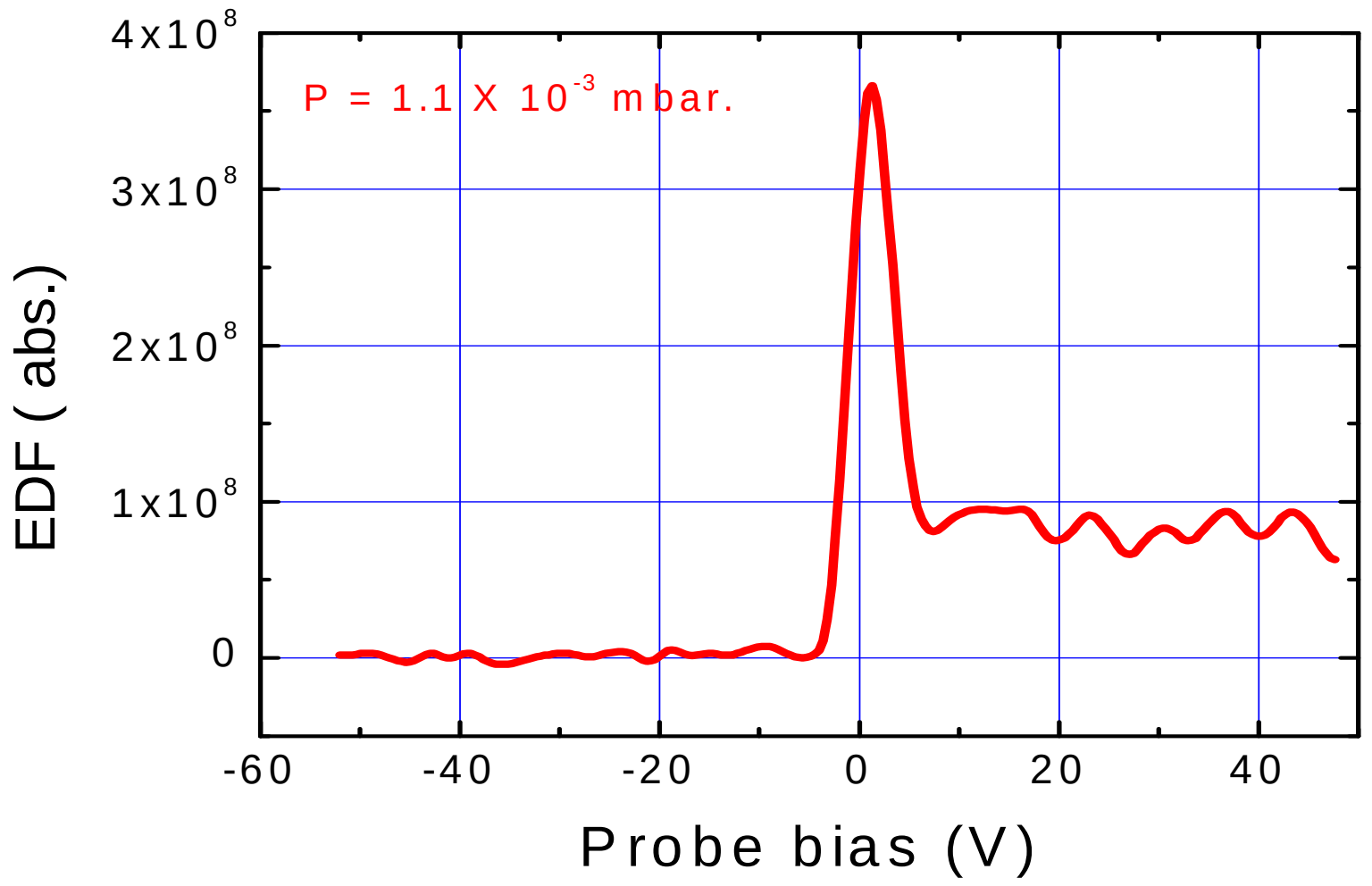

**Fig. 4.** Electron distribution function in arbitrary unit in unperturbed plasma. Gas pressure was $1\times10^{-3}\,mbar$ and plasma density was $3.5\times10^{9}\,cm^{-3}$.

After applying a high voltage positive pulse (Fig. 2(a) and 3(a)) to the exciter, in unperturbed plasma, solitary electron holes are excited for pulse width $\tau_p < 3\tau_i$ and $\tau_p > 3\tau_i$ (Fig. 3(b) and 2(b) respectively) [20, 21]. The width of SEHs is function of applied external electrostatic potential or field [30] and the damping of SEHs depends upon, mainly, on the interaction of trapped electrons with neutral gas atoms i.e. pressure dependent [12].

In the first case, we have measured the transient EDF for a longer pulse width ($\tau_p > 3\tau_i$), i.e. positive pulse of 1020 V and pulse width 10 $\mu s$ (Fig. 2(a)). Figure 5 shows the temporal current variation of a planar Langmuir probe placed at $d$ = 4 cm away from the exciter. Pressure of gas was $1\times10^{-3}\,mbar$. Current variation profile for a given bias shows that after ending of pulse, current increases up to a certain value, after that it decreases toward zero value.

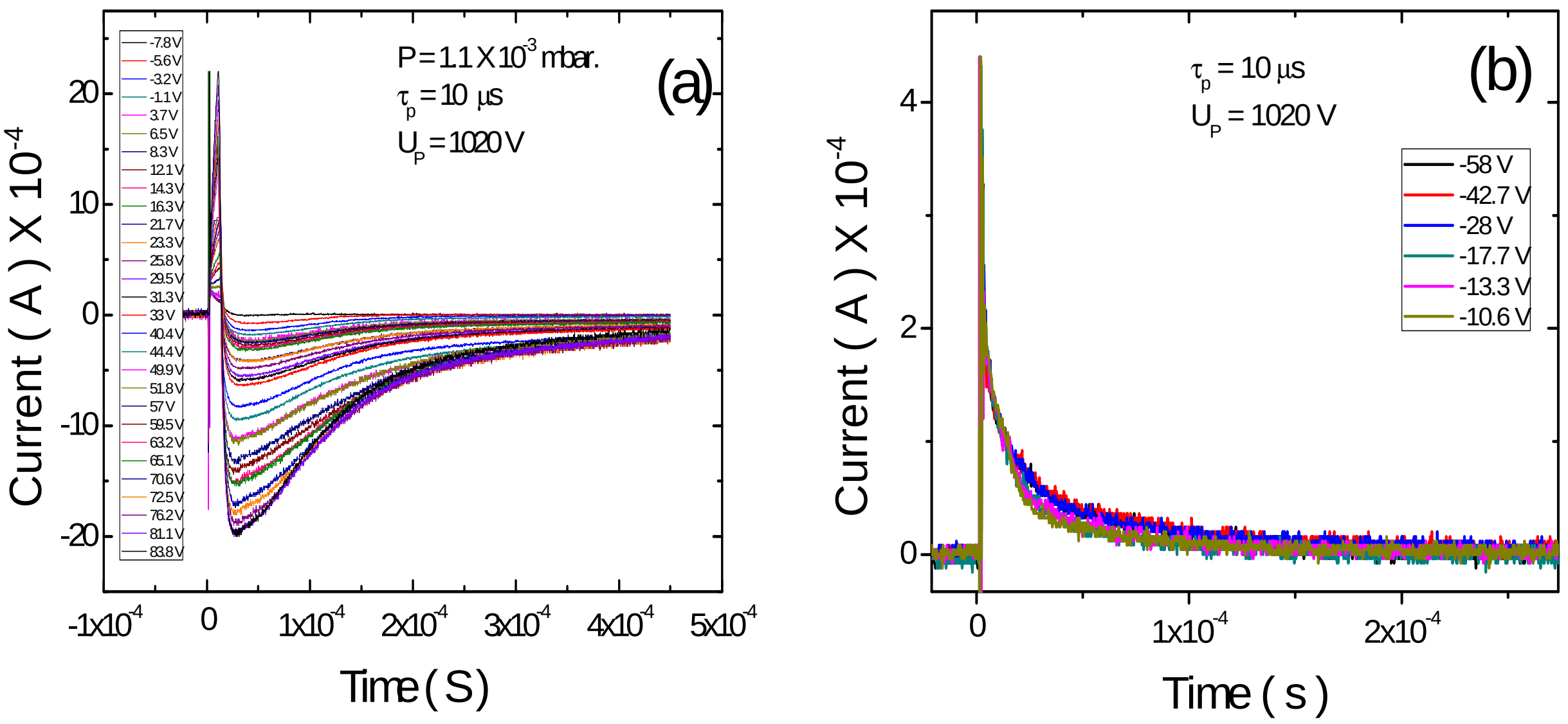

**Fig. 5. (a)** Electron current variation for the positive Langmuir probe bias. **(b)** Ion current variations for negative Langmuir probe bias.

Figure 6(a) shows the perturbed *I-V* characteristics for various times at 4 cm away from the exciter, which is deduced from Fig. 5. These *I–V* characteristics showing a double slope which may be due to the presence of high energy beam electrons. Smoothed curves of Fig. 6 (a) are shown in Fig. 6(b), which are obtained by nonlinear fitting and spline smoothed methods. We used this method for smoothing *I–V* curves before derivative. The standard errors are nearly 5- 10 % in this method.

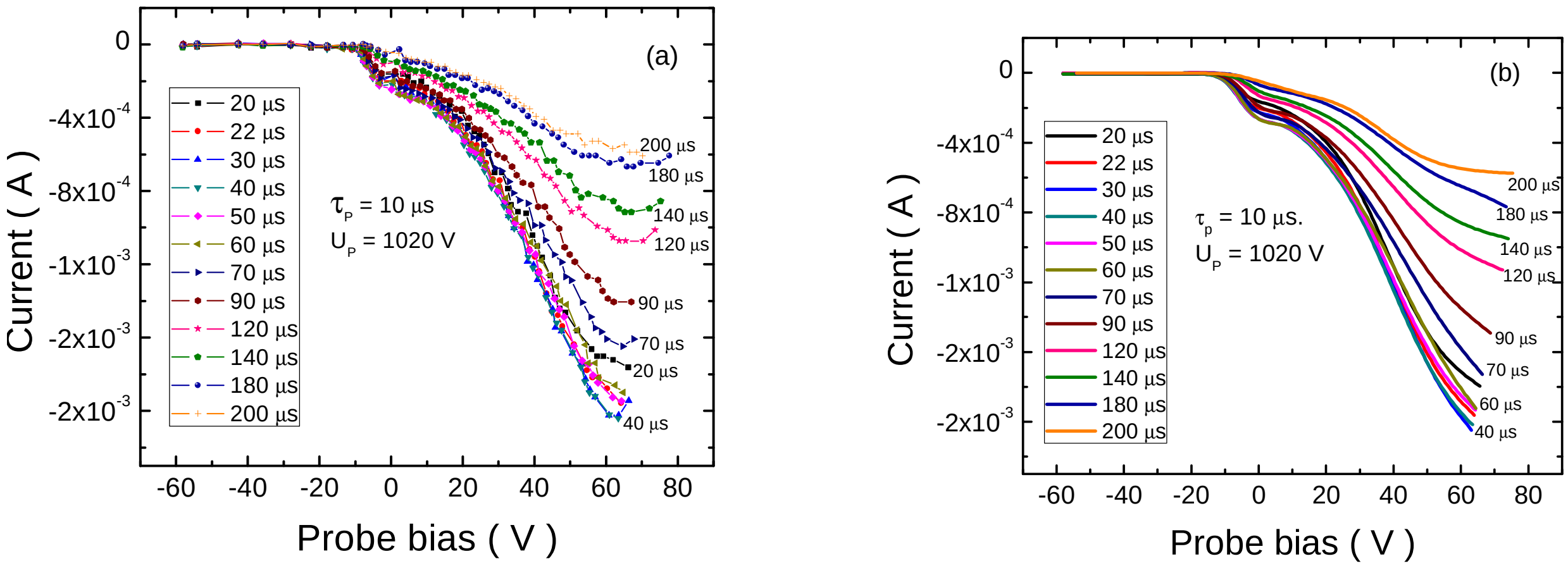

**Fig. 6. (a)** *I-V* characteristics for various times at d = 4 cm. **(b)** Nonlinear fitted and smoothed characteristics curves (of Fig. (a)) for different times of pulse.

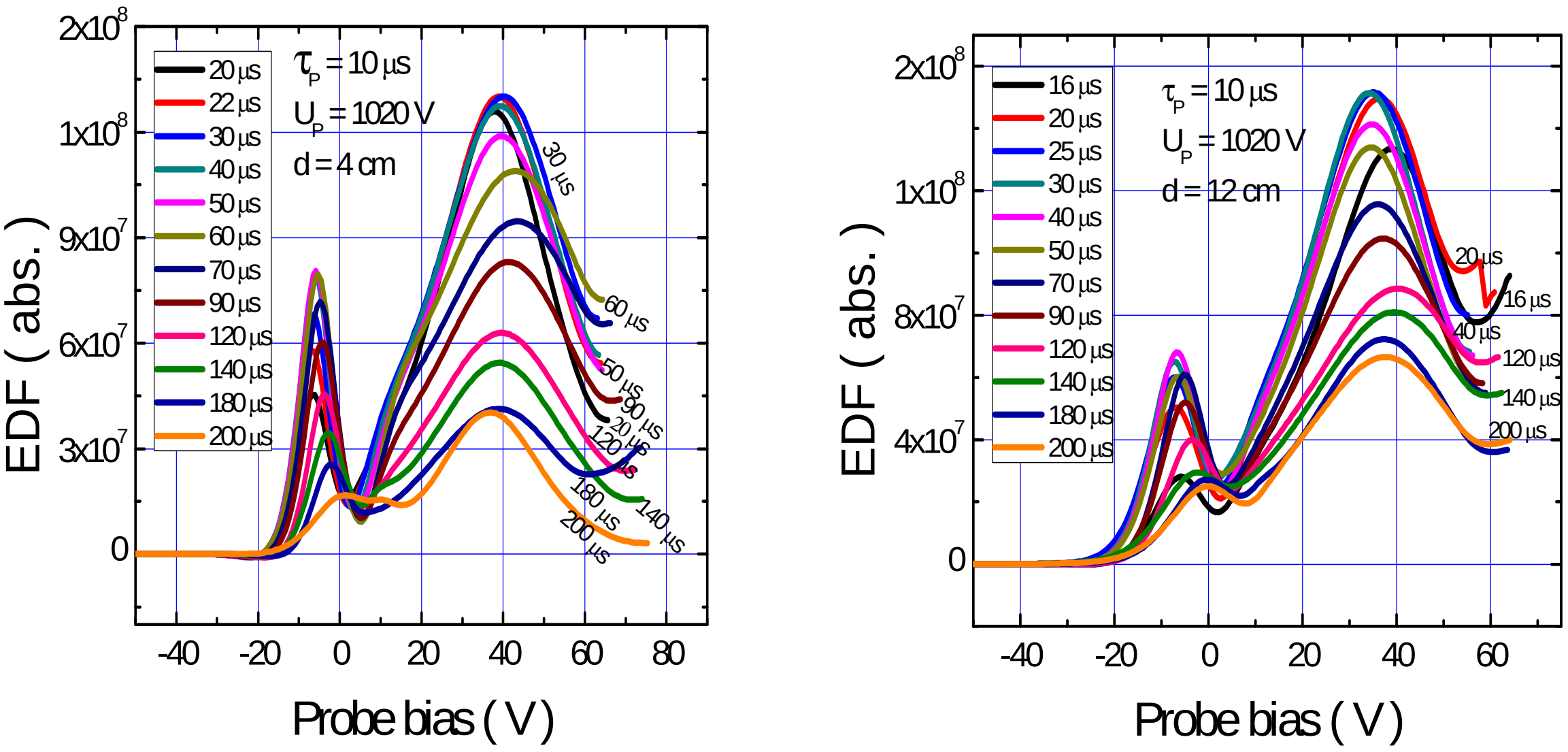

**Fig. 7. (a)** Transient EDF measured at d = 4 cm. **(b)** d = 12 cm for pulse width 10 $\boldsymbol{\mu s}$ and pulse amplitude (U) = 1020 V.

The first derivative of the *I-V* characteristics (of Fig. 6b) by inverting the vertical scale after multiplying the constant factor, represents the transient EDF at a given time. Figure 7 shows the transient EDF for pulse width $\tau > 3\tau_i$ (here 10 $\mu s$) for two different positions (4 cm and 12 cm) away from exciter. This transient EDF shows a double hump profile. The first hump (in the negative bias region) having the high energy electrons and termed as beam or free electrons. Whereas the second hump (in the positive bias region) having the comparatively low energy electrons, which is termed as bulk or trapped electrons. This double hump like profiles are arising after nearly 20 $\mu$s after the application of pulse. Between the time 40 -50 $\mu$s; beam component as well as trapped component have its maximum value and decreases with time evolution. The amplitude and area of the first hump is less than the second hump. So we can say that the population of free electrons are less than the population of trapped electrons. It is seen that the energy of the trapped electrons are approximately same for various times, but the energy of the free electrons are more at the earlier times and decreases as time goes on. The energy of trapped electrons is more at near to the exciter, i.e., 40 V at 4 cm and 35 V at 12 cm. The presence of both components (beam and trapped) in the plasma gives the solution of electron hole. The time around $t \approx 5\times10^{-4} s$ , the transient part of EDF vanishes or SEHs decay and plasma comes in its initial unperturbed state.

Now, we have measured the transient EDF for shorter pulse width ($\tau_p < 3\tau_i$). By changing the capacitance in pulse forming circuit we can change the pulse width. Here the pulse width is 900 ns i.e., $\tau_p < 3\tau_i$ , applied pulse voltage is 1020 V and the Langmuir probe was placed at d = 3 cm away from exciter.

Variation of electron and ion currents with probe bias at a particular time of pulse application to the exciter is given in Fig. 8 (a). Figure 8 (b) is showing the smooth characteristics curves. In the shorter pulse width case, value of the perturbed current (Fig. 8a) is lower than longer pulse width (Fig. 6a), as the other plasma parameters are same.

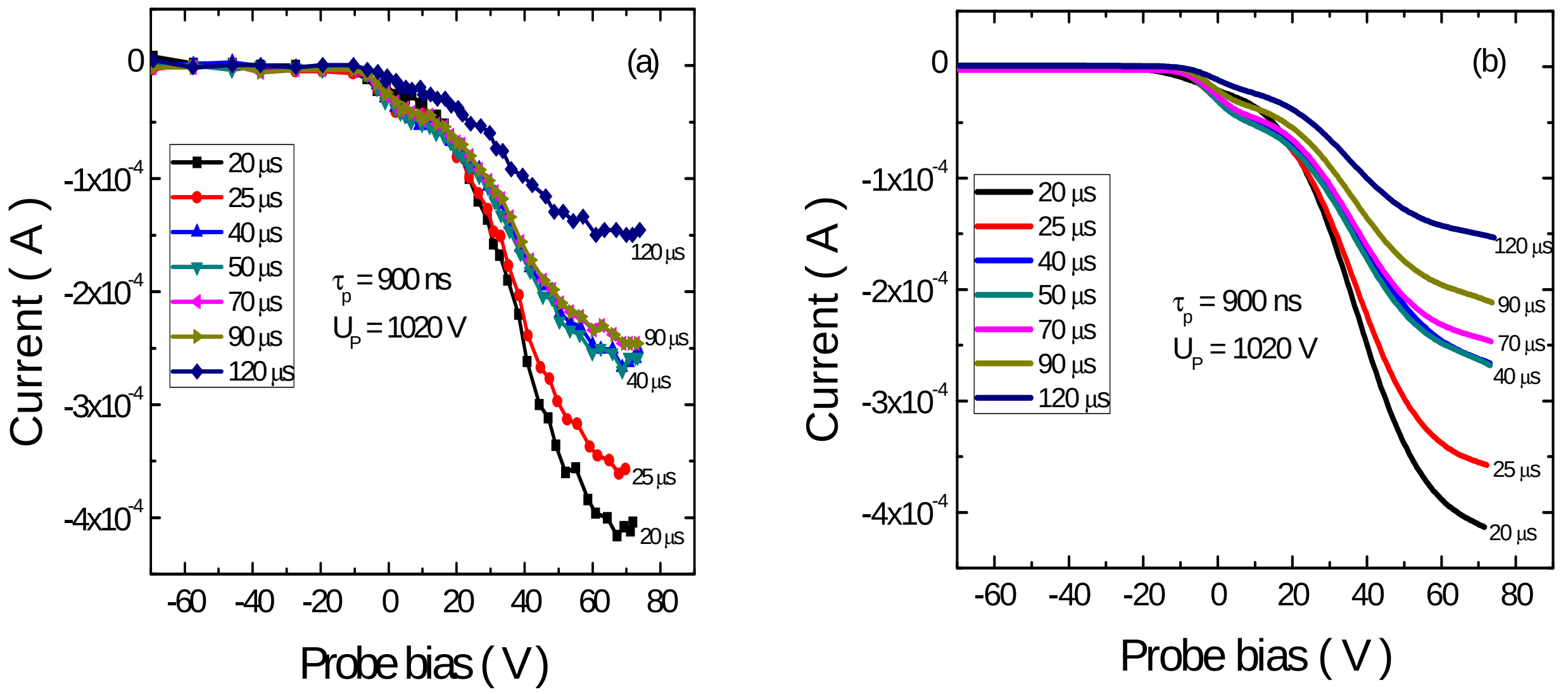

**Fig. 8. (a)** Variation of *I*-*V* characteristics for various times. **(b)** Smoothed curves of Fig. 8 (a). Here, d = 3 cm and pulse width 900 ns

Figure 9 shows the transient EDF at d = 3 cm away from the exciter. About nearly 15$\mu s$, double hump like structures are observed. The SEHs are excited earlier for shorter pulse widths than the longer pulse width (Fig. 7a). Both beam and trapped electron populations are maximize at nearly 20-30 $\mu s$ and after that perturbed part vanishes. Such type of nature is also seen for far away from the exciter. In the small pulse width case, the populations of both components are less compare to larger pulse width (Fig. 7a). It shows the size of SEHs is dependent on pulse width of applied pulse.

The affect of exciter voltage on transient EDF is given in Fig 10. The transient EDF is measured at d = 4 cm away from exciter at a given time t = 40 $\mu s$ for longer applied pulse width. The results show that the population of both components (beam and trapped) and the energy of trapped electrons are increased with increase in excitation potential.

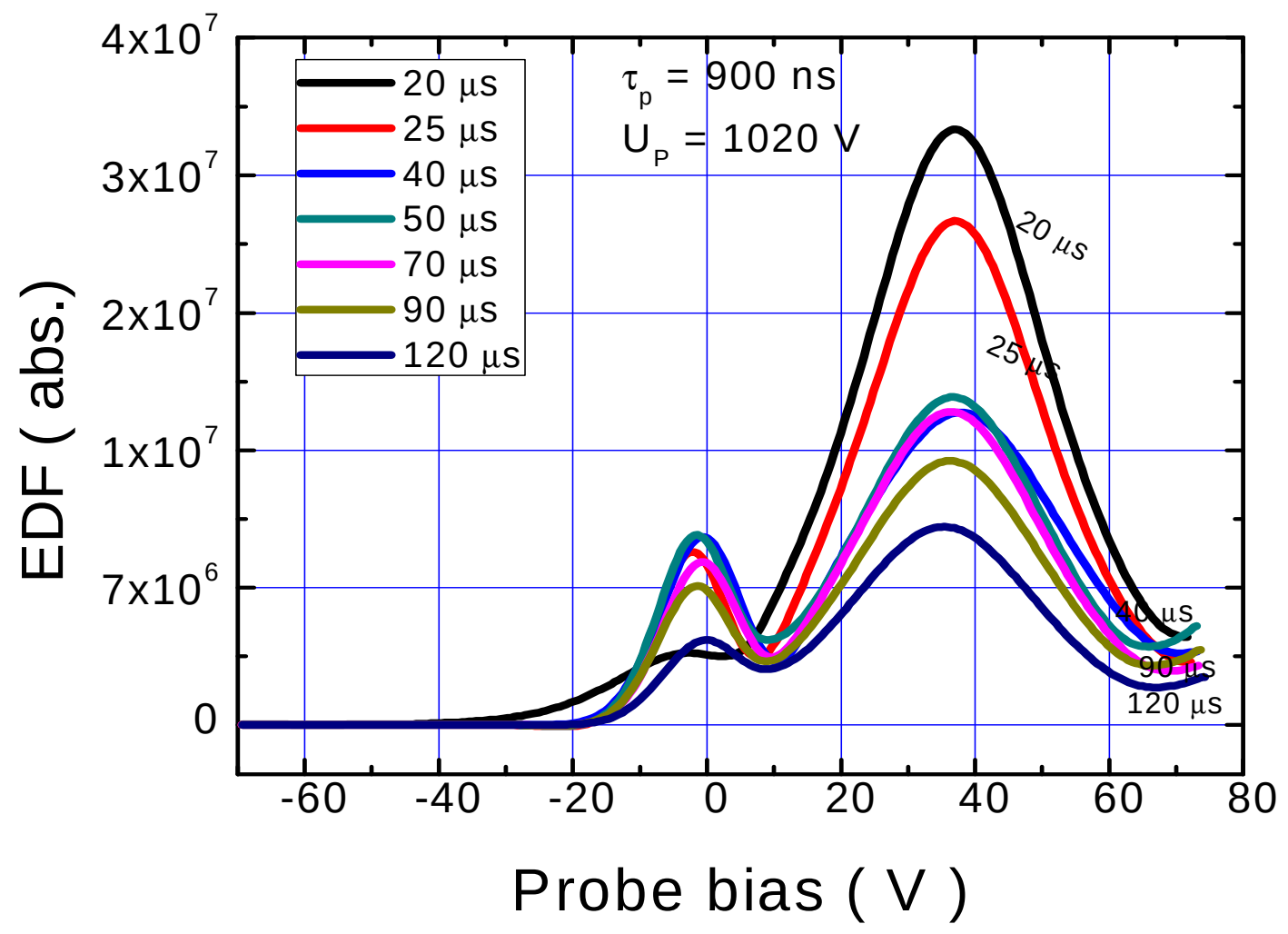

**Fig. 9**. Transient EDF at 3 cm, pulse width 900 ns, $U_P$ = 1020 V.

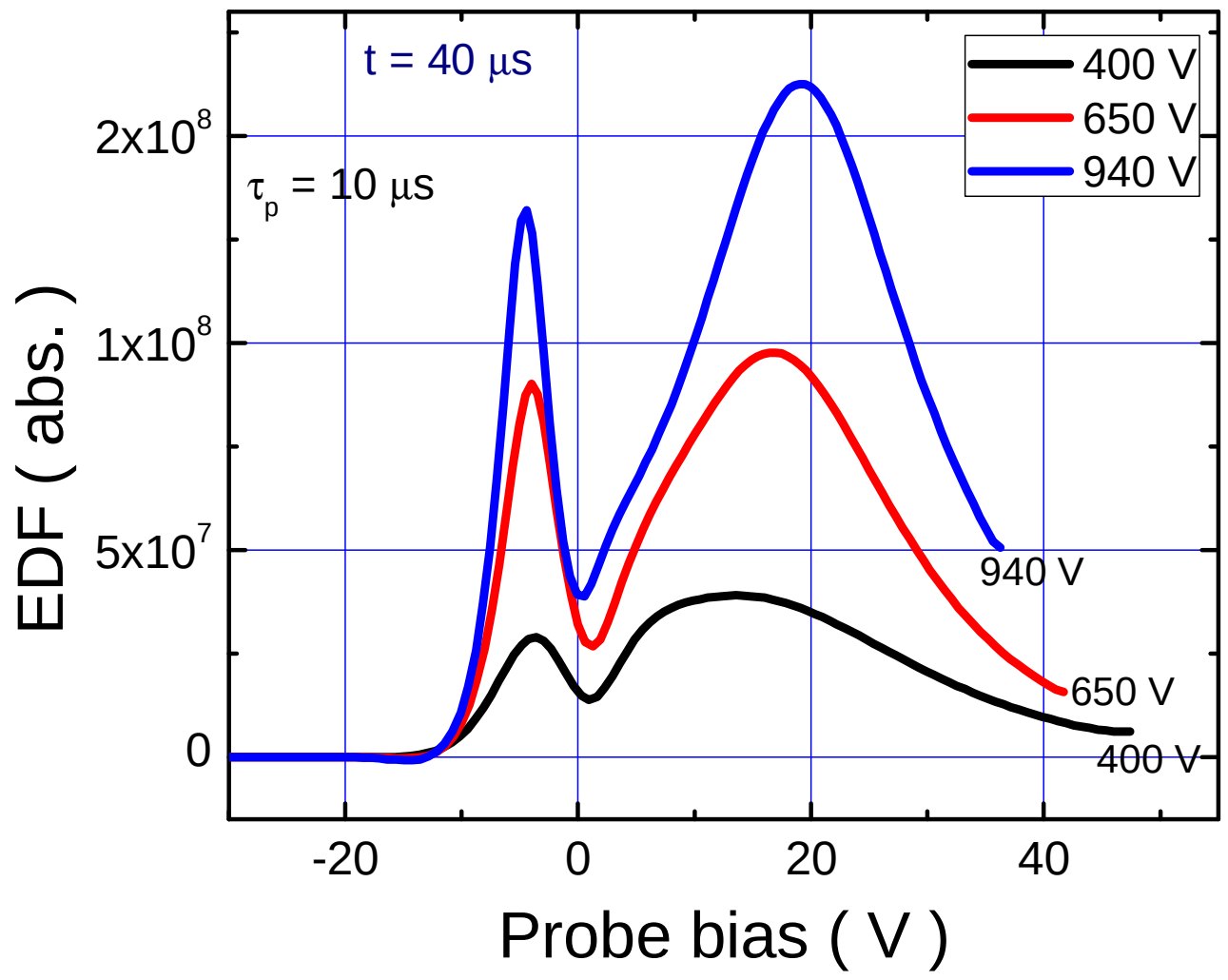

**Fig. 10.** Transient EDF for different exciter voltage at d = 4 cm and applied pulse width $\tau_p$ = 10 $\mu s$, measured at t = 40 $\mu s$ for various applying the pulse amplitude.

The effect of gas pressure on the transient EDF is discussed in this section. With increasing the gas pressure, electron–neutral collision mean free path decreases and probability of collisions increases. Thus, the trajectories of beam and trapped electrons can get modified with electron – neutral collisions. This implies that the SEHs can get destroy with increasing pressure. Figure 11(a) is showing the transient EDF at d = 4 cm from exciter when pulse width was $10\mu s$ and gas pressure was $1\times10^{-3}\,mbar$. Discharge current was .05A. The plasma density was $\approx 3\times10^{9}\,cm.^{-3}$. It’s having a double hump profile. Fig. 11 (b) shows the transient EDF at d = 4 cm from exciter, when the pulse width was 10 $\mu$s, gas pressure was $1.8\times10^{-3}\,mbar$ (higher pressure) and discharge current was .06A.

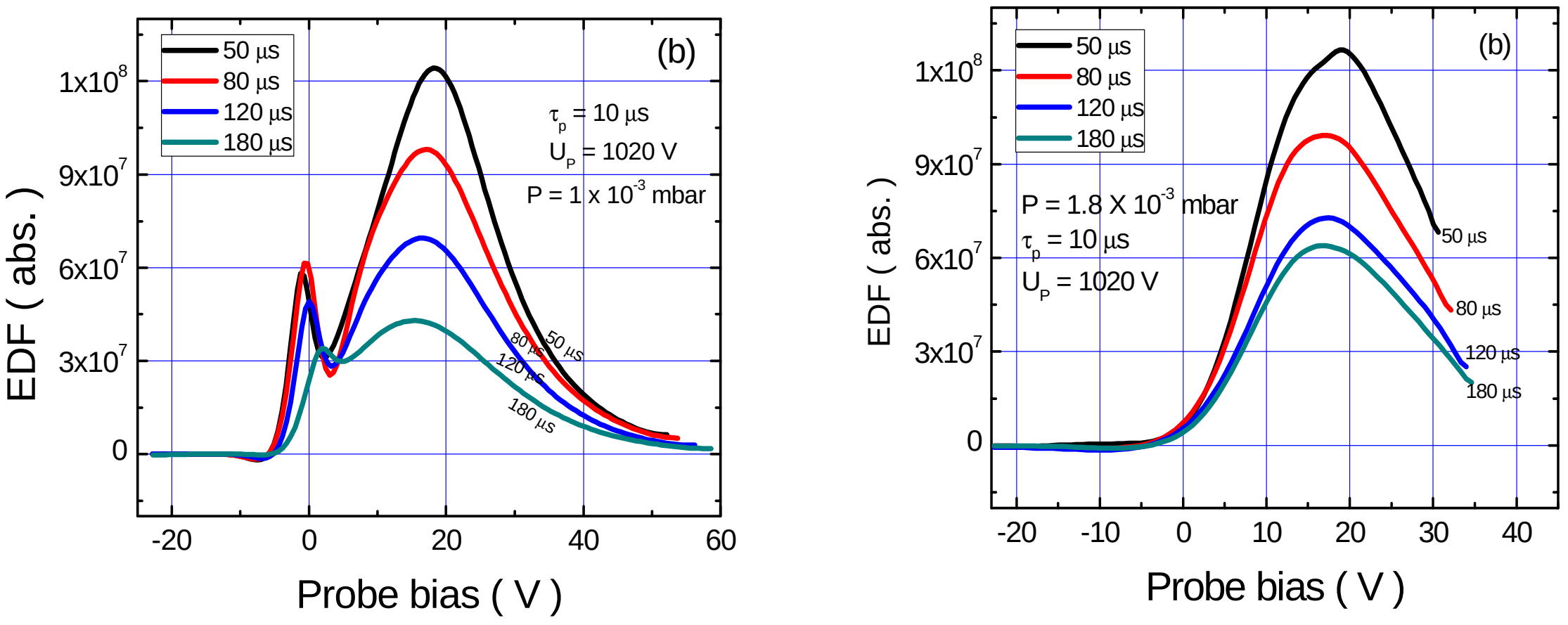

**Fig.11. (a)** Temporal perturbed EDF at d = 4 cm for different times, when pressure P = $1\times10^{-3}$ mbar. **(b)** Transient EDF for different times at d = 4 cm, pressure P = $1.8\times10^{-3}$ mbar.

In higher gas pressure case, beam component is not seen in the transient EDF. Only bulk electron population exists in the plasma, which does not support the existence of SEHs. These results are in favor of collisional effects on the solitary electron hole i.e. solitary electron holes get destroyed due to electron–neutral collisions. This observation is reflected in Ref. [12], where increase in pressure leads to destruction of SEHs.

SEH is excited in laboratory plasma by fast rising high voltage pulses of pulse widths $\tau_p < 3\tau_i$ and $\tau_p > 3\tau_i$. These structures propagate in the order of electron thermal speed [20, 21]. The presence of SEHs in the plasma can be predicted by profile of transient electron distribution function. Electrons obey nearly Maxwellian in unperturbed state; therefore EDF is single hump. Double hump like profile of transient EDF is an evidence for existence of SEHs in the system. Such profile consists of two types population: free (beam) electrons and trapped (bulk) electrons.

All experimental results confirm the existence of SEHs in the plasma for longer time after the pulse is withdrawn. Since, plasma is collisionless therefore these structures are long lived. Collective interaction between electron species, electron–neutrals, electron–ion may be responsible for the destruction of SEHs, as a result, system will turn towards the unperturbed state.

## IV. CONCLUSIONS

The work presented in this report is related to the temporal evolution of electron distribution function after formation of solitary electron holes in unmagnetized, low pressure plasma. The high energy free (or beam type) electrons appear along with the low energy trapped electrons during the plasma evolution after the end of the external pulse. i.e., double hump like profile of transient EDF. Double hump profile of transient EDF gives the solution of SEHs in low pressure plasma. For higher gas pressures, these structures do not exist. In our experiments, it is seen that collective interaction of plasma species will turn the system to unperturbed state after long time. It means that the SEHs decay via collective interaction of plasma species.